\documentclass[aps,prl,twocolumn,numerical,superscriptaddress,nofootinbib,showpacs,showkeys,longbibliography]{revtex4-1}

\usepackage{graphicx,color}
\usepackage{amsmath,amssymb,amsfonts}
\usepackage[colorlinks=true,linkcolor=blue,plainpages=false]{hyperref}
\usepackage{url}

\newcommand{\be}{\begin{equation}}
\newcommand{\ee}{\end{equation}}
\newcommand{\ba}{\begin{eqnarray}}
\newcommand{\ea}{\end{eqnarray}}
\newcommand{\ban}{\begin{eqnarray*}}
\newcommand{\ean}{\end{eqnarray*}}

\def\v2{\mbox{$v_2$}}

\def\sqrtsNN{\mbox{$\sqrt{s_{\mathrm{NN}}}$}}

\begin{document}

\title{Examination of the observability of a chiral magnetically-driven charge-separation difference in
collisions of the $\mathrm{^{96}_{44}Ru +\, ^{96}_{44}Ru}$ and $\mathrm{^{96}_{40}Zr +\, ^{96}_{40}Zr}$ 
isobars at energies \\ available at the BNL Relativistic Heavy Ion Collider}
%\medskip

\author{Niseem~Magdy} 
%\email{niseemm@gmail.com}
\affiliation{Department of Chemistry, Stony Brook University, Stony Brook, New York 11794, USA}

\author{Shuzhe Shi}
%\email{shishuz@umail.iu.edu}
\affiliation{Physics Department and Center for Exploration of Energy and Matter,
Indiana University, 2401 N Milo B. Sampson Lane, Bloomington, IN 47408, USA.}

\author {Jinfeng Liao} 
\email{liaoji@indiana.edu}
\affiliation{Physics Department and Center for Exploration of Energy and Matter,
Indiana University, 2401 N Milo B. Sampson Lane, Bloomington, IN 47408, USA.}

\author {Peifeng Liu} 
\affiliation{Department of Chemistry, Stony Brook University, Stony Brook, New York 11794, USA}
\affiliation{Department of Physics, Stony Brook University, Stony Brook, New York 11794, USA}

\author{Roy~A.~Lacey} 
\email{Roy.Lacey@stonybrook.edu}
\affiliation{Department of Chemistry, Stony Brook University, Stony Brook, New York 11794, USA}
\affiliation{Department of Physics, Stony Brook University, Stony Brook, New York 11794, USA}

\date{\today}

\begin{abstract}
 Anomalous Viscous Fluid Dynamics (AVFD) model calculations for $\mathrm{^{96}_{44}Ru +\, ^{96}_{44}Ru}$
and $\mathrm{^{96}_{40}Zr +\, ^{96}_{40}Zr}$ collisions ($\sqrtsNN~=~200$ GeV) are used in concert with a 
charge-sensitive correlator, to test its ability to detect and characterize the 
charge separation difference expected from the Chiral Magnetic Effect (CME) in these 
isobaric collisions. The tests indicate a larger charge separation for $\mathrm{^{96}_{44}Ru +\, ^{96}_{44}Ru}$ 
than for $\mathrm{^{96}_{40}Zr +\, ^{96}_{40}Zr}$ collisions, and a discernible CME-driven 
difference of $\sim 10$\% in the presence of realistic non-CME backgrounds. 
They also indicate a strategy for evaluating the relative influence of the 
background correlations, present for each isobar.
These results suggest that charge separation measurements for these isobaric species, 
could serve to further constrain unambiguous identification and characterization of the CME in 
upcoming measurements at RHIC.  	
\end{abstract}

\pacs{25.75.-q, 25.75.Gz, 25.75.Ld}% PACS, the Physics and Astronomy
%                             %display desired
\maketitle

%=======================================
% Introduction
%=====================================
Isobaric collisions of $\mathrm{^{96}_{44}Ru +\, ^{96}_{44}Ru}$
and $\mathrm{^{96}_{40}Zr +\, ^{96}_{40}Zr}$ at $\sqrtsNN~=~200$ GeV, will 
be used in an upcoming experiment at the Relativistic Heavy Ion Collider (RHIC)
to measure and characterize a possible charge separation difference, induced 
by the Chiral Magnetic Effect (CME) in these collisions \cite{Kharzeev:2015znc,Skokov:2016yrj}. 
Experimental validation of this purported signal, would constitute an 
invaluable constraint to further identify and characterize the CME in heavy ion collisions.
In addition, it could provide crucial insight on anomalous transport and the interplay 
of chiral symmetry restoration, axial anomaly, and gluonic topology in the QGP.

The rationale for the isobaric collision experiment, stems from two invaluable considerations.
The first is that, by choosing isobars, the well known background correlations 
which complicate charge separation measurements 
\cite{Wang:2009kd,Bzdak:2010fd,Schlichting:2010qia,Muller:2010jd,Liao:2010nv,Khachatryan:2016got}
are made similar, and this greatly facilitates a comparison of the measurements 
for both systems. The second is the expectation that the net axial-charge asymmetry of 
the chiral quarks in the quark-gluon plasma (QGP) \cite{Moore:2010jd,Mace:2016svc} created 
in the isobaric collisions, are similar, but the time-dependent electromagnetic $\vec{B}$ 
fields \cite{Skokov:2009qp,McLerran:2013hla,Tuchin:2014iua} produced for similar impact parameter, 
is larger for $\mathrm{^{96}_{44}Ru +\, ^{96}_{44}Ru}$, due to its larger charge to mass ratio.  
Thus, the chiral anomaly for $\mathrm{^{96}_{44}Ru +\, ^{96}_{44}Ru}$ should convert into a 
stronger electric current than for $\mathrm{^{96}_{40}Zr +\, ^{96}_{40}Zr}$,
\begin{eqnarray} \label{eq_cme}
\vec{J}_Q &=& \sigma_5 \vec{B},\ \ \ \ \sigma_5 = \mu_5 \frac{Q^2}{4\pi^2},
\end{eqnarray} 
leading to a larger final-state charge separation perpendicular to the reaction plane ($\mathrm{\Psi_{RP}}$) 
defined by the impact parameter and the beam axis 
\cite{Kharzeev:2004ey,Kharzeev:2007tn,Kharzeev:2007jp,Fukushima:2008xe,Kharzeev:2010gr,Jiang:2016wve,Skokov:2016yrj}; 
%For recent reviews, see e.g. \cite{Kharzeev:2013ffa,Liao:2014ava,Kharzeev:2015znc}.
here, $\sigma_5$ is the chiral magnetic conductivity, 
$\mu_5$ is the chiral chemical potential that quantifies the axial charge 
asymmetry or imbalance between right-handed and 
left-handed quarks in the plasma, and $Q$ is the  quark electric 
charge \cite{Fukushima:2008xe,Son:2009tf,Zakharov:2012vv,Fukushima:2012vr}. 

\begin{figure}[t]
%
%\centering
\includegraphics[width=0.80\linewidth, angle=-0]{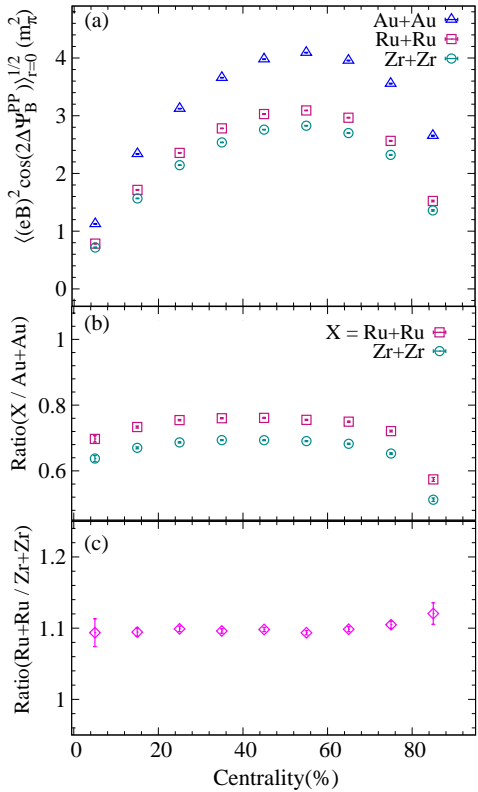} \\
%\includegraphics[width=0.72\linewidth, angle=-0]{eB_Ratio.eps}
%\subfloat{\includegraphics[width=0.70\linewidth, angle=-0]{eB.eps}} \\
%\subfloat{\includegraphics[width=0.70\linewidth, angle=-0]{eB_Ratio.eps}}
%\vskip -0.050in
\caption{(a) Comparison of the centrality dependence of the peak magnetic 
fields ${B}_0$ (perpendicular to the respective participant planes $\mathrm{\Psi^{PP}_{B}}$
which fluctuate about $\mathrm{\Psi_{RP}}$),
for Au+Au Ru+Ru and Zr+Zr collisions at $\sqrtsNN~=~200$ GeV. Panels (b) and (c) show the 
the centrality dependence of the ratio of these peak magnetic fields \cite{isobar-Bfield}.
}
\label{fig1} 
%\vspace{-0.3in}
\end{figure}

Figure~\ref{fig1}(a) shows a clear hierarchy in the magnitude of the peak magnetic 
field values ${B}_0 = \mathrm{\left< (eB)^2\cos(2\Delta \Psi^{PP}_{B}) \right>^{1/2}}$,
evaluated perpendicular to the respective participant planes $\mathrm{\Psi^{PP}_B}$, in  
Au+Au, Ru+Ru  and Zr+Zr collisions at $\sqrtsNN~=~200$ GeV; $\mathrm{\Delta \Psi^{PP}_{B}}$ is the 
angle of the participant plane relative to $\mathrm{\Psi_{RP}}$, so the ${B}_0$ values 
take into account the $\vec{B}$-field fluctuations. The procedure employed to compute ${B_0}$ is 
akin to that employed in Ref.~\cite{Chatterjee:2014sea}.
Figs.\ref{fig1}(b) and (c) show that the ratio of these peak values  
scale approximately as the charge to mass ratio, and the magnetic  
field for $\mathrm{^{96}_{44}Ru +\, ^{96}_{44}Ru}$ 
collisions is $\sim 8-10$\% larger \cite{isobar-Bfield} than that for 
$\mathrm{^{96}_{40}Zr +\, ^{96}_{40}Zr}$ collisions for the same centrality selection. 
%
% Fig.1 
%
Therefore, a major objective of the isobaric collision experiment, is to identify 
and quantify a CME-driven charge separation difference between 
$\mathrm{^{96}_{44}Ru +\, ^{96}_{44}Ru}$ and $\mathrm{^{96}_{40}Zr +\, ^{96}_{40}Zr}$
of order $10$\%, via measurements of the first $P$-odd sine term ($\mathrm{a_{1}}$) in 
the Fourier decomposition of the charged-particle azimuthal distribution~\cite{Voloshin:2004vk}
for both systems:
\begin{eqnarray}\label{eq:1}
\mathrm{\frac{dN^{ch}}{d\phi} \propto [1 + 2\sum_{n} v_{n} \cos(n \Delta\phi) + a_n sin(n \Delta\phi)  + ...]}
\end{eqnarray}
where $\mathrm{\Delta\phi = \phi -\Psi_{RP}}$ gives the particle azimuthal angle
with respect to the reaction plane angle, and $\mathrm{v_{n}}$ and $\mathrm{a_{n}}$ denote the
coefficients of $P$-even and $P$-odd Fourier terms, respectively. 
The second-order event plane, $\Psi_{2}$, determined by the maximal particle density 
in the elliptic azimuthal anisotropy and the beam axis, serves as a proxy for $\mathrm{\Psi_{RP}}$
in experimental measurements. 

In addition to the ${B}_0$ difference indicated for the isobars, 
it is noteworthy that Fig.~\ref{fig1} also suggest a specific hierarchy, as well as 
patterns in the relative magnitudes for CME-driven charge separation for the systems indicated.
The observation of such magnitudes and trends in future charge separation measurements could 
also serve as an important constraint.

A caveat on the proposed measurement for the isobars, is that the initial axial charge 
and the time evolution of the magnetic field (c.f. Eq.~\ref{eq_cme}) are 
unconstrained theoretically. Thus, it is not 
certain whether the expected $\vec{B}$-driven charge separation difference would 
remain detectable, after possible signal losses associated with the dynamics of the 
evolution from the QGP phase to particle freeze-out. 
It is also uncertain whether a charge separation difference that survives the 
reaction dynamics, would still be discernible in the presence of the well known 
background correlations which contribute and complicate the measurement of CME-driven 
charge separation
\cite{Wang:2009kd,Bzdak:2010fd,Schlichting:2010qia,Muller:2010jd,Liao:2010nv,Khachatryan:2016got}.

In recent work \cite{Magdy:2017yje}, we have developed and tested a new correlator 
designed to give discernible responses for background- and CME-driven charge 
separation relative to the $\Psi_{3}$ and $\Psi_{2}$ event planes respectively. 
An initial rudimentary comparison of the correlators for preliminary 
data \cite{Laceytalk2017} and theory, indicated results compatible with 
a CME-driven charge separation in 40-50\% central Au+Au collisions at $\sqrtsNN~=~200$ GeV.
Here, we use this correlator in concert with state-of-the-art 
Anomalous Viscous Fluid Dynamics (AVFD) model calculations, to test its efficacy to detect 
and characterize the expected charge separation difference, induced by the Chiral Magnetic Effect (CME), 
in $\mathrm{^{96}_{44}Ru +\, ^{96}_{44}Ru}$ and $\mathrm{^{96}_{40}Zr +\, ^{96}_{40}Zr}$ 
collisions. The tests are performed under ``realistic'' AVFD model conditions, which mimic the complications 
that could result from the influence of both the reaction dynamics and background correlations.

\begin{figure}[t]
\includegraphics[width=0.70\linewidth, angle=-90]{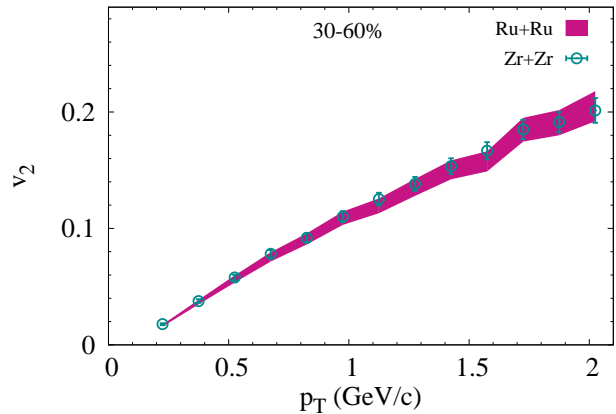}
\vskip -0.10in
\caption{ Comparison of the differential elliptic flow $\mathrm{v_2(p_T)}$, 
extracted from AVFD events for 30-60\% central $\mathrm{^{96}_{44}Ru +\, ^{96}_{44}Ru}$ 
and $\mathrm{^{96}_{40}Zr +\, ^{96}_{40}Zr}$ collisions at $\sqrtsNN~=~200$ GeV. 
The extractions were carried out for charged hadrons with $|\eta| \alt 1.5$ with the event
planes constructed in the range $2.5\alt |\eta| \alt 4.0$.
} 
\label{fig2} 
%\vspace{-0.1in}
\end{figure}

The AVFD model \cite{Jiang:2016wve,Shi:2017cpu} uses Monte Carlo Glauber initial conditions to simulate the evolution 
of fermion currents in the QGP, on top of the bulk fluid evolution implemented in 
the VISHNU hydrodynamic code \cite{Song:2010mg}, supplemented with a URQMD hadron 
cascade ``afterburner'' stage. 
The code gives a good representation of the experimentally measured bulk properties -- 
particle yields, spectra, flow, etc. Therefore, it gives a realistic estimate of the 
magnitude and trends of the background correlations expected in the isobaric data samples 
to be obtained at RHIC. Fig.~\ref{fig2} shows that the AVFD values for 
elliptic flow ($\mathrm{v_2(p_T)}$), a major driver of background correlations, is similar 
for the two isobars and is in line with expectation that the background correlations for 
$\mathrm{^{96}_{44}Ru +\, ^{96}_{44}Ru}$ and $\mathrm{^{96}_{40}Zr +\, ^{96}_{40}Zr}$ 
should be similar for the indicated centrality range \cite{Skokov:2016yrj,Deng:2016knn}. 
The two isobars have a small deformation difference which leads to eccentricity ($\varepsilon_2$) 
differences and consequently, a $\mathrm{v_2}$ differences for centralities $\alt 30$\% \cite{Deng:2016knn}.
However, such a difference would not have a strong impact on our analysis as discussed 
below.

%
% Fig.2
%
%%%%%%%%%%%%%%%%%%%%%%%%%%%%%%%
%

Anomalous transport from the CME, is also implemented in the AVFD model. This is accomplished
via a time-dependent magnetic field $B(\tau) = \frac{B_0}{1+\left(\tau / \tau_B\right)^2}$,
which acts in concert with a nonzero initial axial charge density, to generate 
a CME current (encoded in the fluid dynamical equations).
The peak values $B_0$, of the magnetic field, is obtained from event-by-event 
simulations~\cite{Bloczynski:2012en}, and are used with a relatively conservative 
lifetime $\tau_B=0.6$ fm/c, to generate the time-dependence of the magnetic field.
The estimate for the initial axial charge density, which results from gluonic topological charge fluctuations, 
is based on the strong chromo-electromagnetic fields developed in the 
early-stage glasma phase \cite{Kharzeev:2001ev,Mace:2016svc}.

With these essential ingredients, the AVFD model was used to simulate a charge separation along 
the $\vec{B}$-field (i.e. perpendicular to the reaction plane), in combination with background correlations, 
for $\mathrm{^{96}_{44}Ru +\, ^{96}_{44}Ru}$
and $\mathrm{^{96}_{40}Zr +\, ^{96}_{40}Zr}$ collisions at $\sqrtsNN~=~200$ GeV.
A subsequent analysis of these simulated events was performed to identify 
and quantify the predicted CME-driven charge separation difference between 
the two isobars.
%
% Fig.3
%
\begin{figure*}[t]
\includegraphics[width=0.5\linewidth, angle=-90]{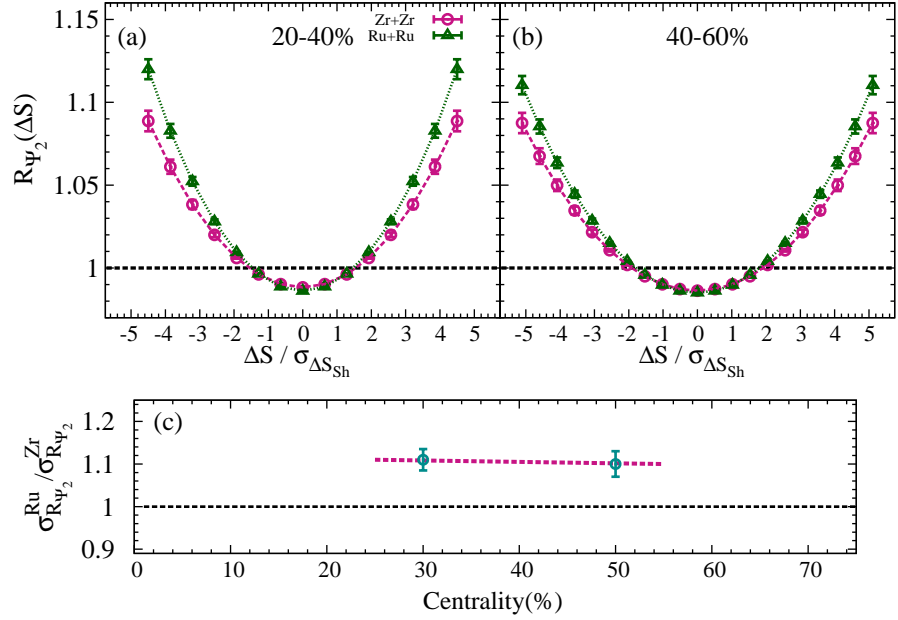}
\vskip -0.10in
\caption{ Comparison of the $R_{\Psi_2}(\Delta S)$ correlators obtained from 
simulated AVFD events for 
$\mathrm{^{96}_{40}Zr +\, ^{96}_{40}Zr}$ and $\mathrm{^{96}_{44}Ru +\, ^{96}_{44}Ru}$ 
at $\sqrtsNN~=~200$ GeV, for $20-40$\% (a) and $40-60$\% (b) central collisions.
Panel (c) shows the ratio $\mathrm{\sigma_{R^{Zr}_{\Psi_2}}}/\mathrm{\sigma_{R^{Ru}_{\Psi_2}}}$, 
of the Gaussian widths extracted from the correlators for the isobars. The magnitude of the 
charge separation is inversely proportional to the width of the $R_{\Psi_2}(\Delta S)$ correlator. 
} 
\label{fig3} 
%\vspace{-0.1in}
\end{figure*} 
%%%%%%%%%%%%%%%%%%%%%%%%%%%%%%%
%
%The correlation function in the numerator of Eq.~\ref{eq:4} is obtained 
%as 
%

The correlator $R_{\Psi_2}(\Delta S)$, constructed relative to the $\Psi_2$ plane, 
was used for the charge separation measurements. As outlined in Ref. \cite{Magdy:2017yje},
the correlators $R_{\Psi_m}(\Delta S)$ can be expressed as the ratios:
\be
R_{\Psi_m}(\Delta S) = C_{\Psi_m}(\Delta S)/C_{\Psi_m}^{\perp}(\Delta S), \, m=2,3 ,
\label{eq:4}
\ee
where $C_{\Psi_m}(\Delta S)$ and $C_{\Psi_m}^{\perp}(\Delta S)$ are correlation functions
designed to quantify charge separation $\Delta S$, parallel and perpendicular (respectively) to 
the $\vec{B}$ field, i.e., perpendicular and parallel (respectively) to $\mathrm{\Psi_{RP}}$. 
The correlation function $C_{\Psi_2}(\Delta S)$ measures 
both CME- and background-driven charge separation, while $C_{\Psi_2}^{\perp}(\Delta S)$ measures 
only background-driven charge separation. The $C_{\Psi_3}(\Delta S)$ and $C_{\Psi_3}^{\perp}(\Delta S)$
correlation functions can also be used to provide insight on the importance 
of background-driven charge separation, because they are both insensitive to a CME-driven charge 
separation \cite{Magdy:2017yje}. However, they are not essential for the present 
analysis, as will be shown below.

The $C_{\Psi_2}(\Delta S)$ correlation function, used to quantify charge separation parallel to the $\vec{B}$ field, 
is constructed from the ratio of two distributions \cite{Ajitanand:2010rc}: 
\be
C_{\Psi_{2}}(\Delta S) = \frac{N_{\text{real}}(\Delta S)}{N_{\text{Shuffled}}(\Delta S)},
\label{eq:5}
\ee
where $N_{\text{real}}(\Delta S)$ is the distribution over events, of charge separation 
relative to the $\Psi_2$ plane in each event:
\be
\Delta S = \frac{{\sum\limits_1^p {\sin (\frac{m}{2}\Delta {\varphi_{m} })} }}{p} - 
\frac{{\sum\limits_1^n {\sin (\frac{m}{2}\Delta {\varphi_{m}  })} }}{n},
\label{eq:7}
\ee
where $n$ and $p$ are the numbers of negatively- and positively charged hadrons in an event, 
$\Delta {\varphi_{2}}= \phi - \Psi_{2}$ and $\phi$ is the 
azimuthal emission angle of the charged hadrons. The $N_{\text{Shuffled}}(\Delta S)$ distribution
was similarly obtained from the same events, following random reassignment (shuffling) of the charge of 
each particle in an event. This procedure ensures identical properties for the 
numerator and the denominator in Eq.~\ref{eq:5}, except for the charge-dependent correlations 
which are of interest.

The correlation function $C_{\Psi_{2}}^{\perp}(\Delta S)$, used to quantify charge separation 
perpendicular to the $\vec{B}$ field, i.e., background-driven charge separation, was constructed 
with the same procedure outlined for $C_{\Psi_{2}}(\Delta S)$, but with $\Psi_{2}$ replaced by $\Psi_{2}+\pi/2$.
This $\pi/2$ rotation of the event plane, suppresses the contributions from 
CME-driven charge separation to this correlation function.

The correlator $R_{\Psi_2}(\Delta S) = C_{\Psi_2}(\Delta S)/C_{\Psi_2}^{\perp}(\Delta S)$, 
gives a measure of the magnitude of charge separation parallel to the $\vec{B}$ field (perpendicular to $\Psi_2$),
relative to that for charge separation perpendicular to the $\vec{B}$ field (parallel to $\Psi_2$).
Consequently, correlations dominated by CME-driven charge separation are expected to result in 
a concave-shaped distributions with widths that are larger for $\mathrm{^{96}_{40}Zr +\, ^{96}_{40}Zr}$
than for $\mathrm{^{96}_{44}Ru +\, ^{96}_{44}Ru}$. That is, the stronger CME-driven charge separation 
expected for $\mathrm{^{96}_{44}Ru +\, ^{96}_{44}Ru}$ collisions, should lead to 
a narrower $R_{\Psi_2}(\Delta S)$ distribution. Here, it is noteworthy that the absence of a strong 
correlation between the orientation of the $\Psi_3$ plane and the $\vec{B}$-field, makes $R_{\Psi_3}(\Delta S)$ 
insensitive to the CME signal, but sensitive to background-driven charge separation. Thus, 
a background-driven charge separation would lead to similar patterns for 
the $R_{\Psi_2}(\Delta S)$ and $R_{\Psi_3}(\Delta S)$ correlators, while a CME-driven charge separation 
would result in characteristically different patterns for the two correlators.

The magnitude of a CME-driven charge separation is reflected in the width of the concave-shaped 
distribution for $R_{\Psi_2}(\Delta S)$ \cite{Magdy:2017yje}, which is also influenced by particle number 
fluctuations and the resolution of $\Psi_2$. 
That is, stronger CME-driven signals lead to narrower concave-shaped distributions (smaller widths), 
which are made broader by particle number fluctuations and poorer event-plane resolutions. 
The influence of the particle number fluctuations can be minimized by scaling $\Delta S$ 
by the width $\mathrm{\sigma_{\Delta_{Sh}}}$ of the distribution 
for $N_{\text{shuffled}}(\Delta S)$ {\em i.e.}, $\Delta S^{'} = \Delta S/\mathrm{\sigma_{\Delta_{Sh}}}$. 
%Similarly, the effects of the event plane resolution can be accounted for by scaling 
%$\Delta S^{'}$ by the resolution factor $\mathrm{\delta_{Res} = e^{0.5(1-Res)^2}}$, {\em i.e.}, 
%$\Delta S^{''}= \Delta S^{'}/\mathrm{\delta_{Res}}$, where $\mathrm{Res}$ is the event plane resolution.
Very little, if any, difference in the event plane resolution for the two isobars, is expected 
for the centrality range discussed. Therefore, it is not necessary to consider the effects of event 
plane resolution in the evaluations to obtain the relative charge separation difference.

Figure \ref{fig3} compares the $R_{\Psi_2}(\Delta S^{'})$ correlators obtained from 
currently available AVFD events for $\mathrm{^{96}_{40}Zr +\, ^{96}_{40}Zr}$
collisions $\mathrm{^{96}_{44}Ru +\, ^{96}_{44}Ru}$ at $\sqrtsNN~=~200$ GeV.
Panels (a) and (b) show the results for 20-40\% and 40-60\% central 
collisions respectively. In each plot, we have scaled $\Delta S$ by the width 
$\mathrm{\sigma_{\Delta S_{Sh}}}$, of the ${N_{\text{Shuffled}}}$ distribution, to 
account for possible differences in the associated number fluctuations due to   
charge particle multiplicity differences for the isobars at each centrality. 
The concave-shaped distributions, apparent in each panel, confirm the AVFD input 
CME-driven signal for the isobars. To quantify the isobaric charge separation difference at 
each centrality, a Gaussian fit (indicated by the curves) was made to the respective 
correlators in each panel, to extract the associated widths.  

The apparent difference in the widths $\mathrm{\sigma_{R_{\Psi_2}}}$, of the distributions 
for $\mathrm{^{96}_{40}Zr +\, ^{96}_{40}Zr}$ and $\mathrm{^{96}_{44}Ru +\, ^{96}_{44}Ru}$
reflects the increase in charge separation expected from the difference in the magnitude 
of the $\vec{B}$ field for the two isobars. Fig. \ref{fig3}(c) shows the
ratio $\mathrm{\sigma_{R^{Zr}_{\Psi_2}}}/\mathrm{\sigma_{R^{Ru}_{\Psi_2}}}$, for the two 
centrality selections. They indicate a charge separation which is $\sim 10$\% larger for 
$\mathrm{^{96}_{44}Ru +\, ^{96}_{44}Ru}$ as expected from the magnetic field difference. 
The observed trend with centrality is also consistent with expectation, c.f. Fig.\ref{fig1}.
The implied sensitivity of these results suggests that the $R_{\Psi_{2}}(\Delta S)$ 
correlator would be able to discern and quantify a CME-driven charge separation 
signal of comparable magnitude, in the upcoming isobaric collisions experiment at RHIC.
Here, it is noteworthy that in a recent study \cite{Sun:2018idn} the commonly employed Gamma 
correlator \cite{Abelev:2009ac,Abelev:2009ad,Adamczyk:2013hsi,Adamczyk:2013kcb,Adamczyk:2014mzf,Tribedy:2017hwn,Zhao:2017nfq}
was shown to be more strongly influenced by background correlations than the $R_{\Psi_{2}}(\Delta S)$
correlator. 

The relative influence of the background correlations for the two isobars can also 
be studied directly, via isobaric ratios of the correlation functions:
\ba
 \frac{[C_{\Psi_{2}}(\Delta S)]_{\mathrm{Ru+Ru}}}{[C_{\Psi_{2}}(\Delta S)]_{\mathrm{Zr+Zr}}},  \quad  \quad 
 \frac{[C_{\Psi_{2}}^{\perp}(\Delta S)]_{\mathrm{Ru+Ru}}}{[C_{\Psi_{2}}^{\perp}(\Delta S)]_{\mathrm{Zr+Zr}}}. 
\ea
Here, the essential notion is that, the isobaric ratio of $C_{\Psi_{2}}(\Delta S)$ should be concave-shaped, 
due to the stronger CME-driven charge separation in $\mathrm{^{96}_{44}Ru +\, ^{96}_{44}Ru}$ collisions.
That is, the stronger CME-driven charge separation for $\mathrm{^{96}_{44}Ru +\, ^{96}_{44}Ru}$ would lead 
to a narrower width for the $C_{\Psi_{2}}(\Delta S)$ distribution for the $\mathrm{^{96}_{44}Ru}$ isobar.
In contrast, the isobaric ratio of $C_{\Psi_{2}}^{\perp}(\Delta S)$, which only measures background-driven 
charge separation, should be essentially flat due to the expected similarity of the background correlations 
for the two isobars. A flat distribution would confirm the cancellation of background contributions to 
the isobaric ratio for $C_{\Psi_{2}}(\Delta S)$. A deviation from this flat expectation would provide 
a benchmark for possible differences in background-driven charge separation for the two isobars. Such differences 
could be crosschecked for collision centralities ($\alt 30$\%)
where background differences are expected for the two isobars \cite{Deng:2016knn}. 
Note as well that such a background influence could be further checked via comparisons of the 
$R_{\Psi_3}(\Delta S)$ correlators, since they are insensitive to CME-driven charge 
separation \cite{Magdy:2017yje}.

\begin{figure}[t]
%
%\vskip 0.3cm
\includegraphics[width=0.75\linewidth, angle=-90]{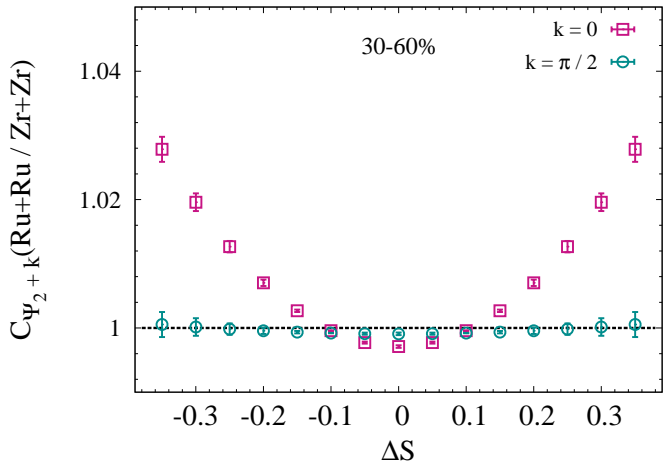}
%\vskip -2.0cm
\caption{ Comparison of the ratio of the $C_{\Psi_2}(\Delta S)$ correlation functions, used to quantify 
charge separation parallel to the $\vec{B}$ field for both isobars, and the ratio of the 
$C_{\Psi_{2}}^{\perp}(\Delta S)$ correlation function, used to quantify charge separation 
perpendicular to the $\vec{B}$ field for both isobars.
} 
\label{fig4} 
\vspace{-0.1in}
\end{figure}

The respective isobaric ratios shown in Fig.~\ref{fig4}, validate the expected patterns for 
the CME- and background-driven signals in AVFD events.  
Note as well that the concavity of the distribution for the $C_{\Psi_{2}}(\Delta S)$ 
isobaric ratio, is consistent with $R_{\Psi_2}(\Delta S)$ correlator 
distributions shown in Fig. \ref{fig3}.
The distributions in Fig. \ref{fig4} constitute an additional cross check for identifying 
and characterizing the CME-driven charge separation difference for the 
$\mathrm{^{96}_{40}Zr +\, ^{96}_{40}Zr}$ and $\mathrm{^{96}_{44}Ru +\, ^{96}_{44}Ru}$
isobars, as well as for quantifying the relative influence of their respective 
background correlations. 

%
% Fig.4
%

In summary, we have used our newly developed $R_{\Psi_{m}}(\Delta S)$ correlator, 
in concert with state-of-the-art Anomalous Viscous Fluid Dynamics (AVFD) model calculations, 
to test its efficacy to detect and characterize a Chiral-Magnetically-induced charge separation
difference in isobaric collisions of $\mathrm{^{96}_{44}Ru +\, ^{96}_{44}Ru}$
and $\mathrm{^{96}_{40}Zr +\, ^{96}_{40}Zr}$ at $\sqrtsNN~=~200$ GeV. The tests indicate a 
discernible CME-driven difference of $\sim 10$\% in the presence of realistic non-CME 
backgrounds. They further indicate that the relative influence of the background 
correlations, present for each isobar, can be quantified.
These results suggest that charge separation measurements for these isobaric species, 
could serve to further constrain unambiguous identification and characterization of both 
CME-driven and background-driven charge separation in heavy ion collisions at RHIC.  	
Preparations are currently underway for experimental running at RHIC, as well as the 
extraction of experimental and theoretical differential $R_{\Psi_{m}}(\Delta S)$ correlators 
for the $\mathrm{^{96}_{44}Ru +\, ^{96}_{44}Ru}$ and $\mathrm{^{96}_{40}Zr +\, ^{96}_{40}Zr}$
isobaric species.

\section*{Acknowledgments}
%=======================================
% Acknowledgements (June/2/2016)
%=======================================
\begin{acknowledgments}
This research is supported by the US Department of Energy, Office of Science, Office of Nuclear Physics, 
under contract DE-FG02-87ER40331.A008  (NM, NA and RL) and by the National Science Foundation 
under Grant No. PHY-1352368 (SS and JL).  
The AVFD study is based upon work supported by the U.S. Department of Energy, 
Office of Science, Office of Nuclear Physics, within the framework of 
the Beam Energy Scan Theory (BEST) Topical Collaboration.
\end{acknowledgments}
%
%%%%%%%%%%%%%%%%%%%%%%%%%%%%%  References  %%%%%%%%%%%%%%%%%%%%%%%%%%%%%%
%
\bibliography{lpvpub} 

\begin{thebibliography}{41}
\bibitem [{iso()}]{isobar-Bfield}%
  \BibitemOpen
  \href@noop {} {}\bibinfo {note} {{The choice of the input model parameters
  for the two isobars, has a modest influence on the magnitude of the
  calculated $\vec{B}$-field difference.}}\BibitemShut {Stop}%
\end{thebibliography}
\end{document}